# Intrusion Detection System: Overview

Hamdan.O.Alanazi, Rafidah Md Noor, B.B Zaidan, A.A Zaidan

**Abstract**-- Network Intrusion Detection (NID) is the process of identifying network activity that can lead to the compromise of a security policy. In this paper, we will look at four intrusion detection approaches, which include ANN or Artificial Neural Network, SOM, Fuzzy Logic and SVM. ANN is one of the oldest systems that have been used for Intrusion Detection System (IDS), which presents supervised learning methods. However, in this research, we also came across SOM or Self Organizing Map, which is an ANN-based system, but applies unsupervised methods. Another approach is Fuzzy Logic (IDS-based), which also applies unsupervised learning methods. Lastly, we will look at the SVM system or Support Vector Machine for IDS. The goal of this paper is to draw an image for hybrid approaches using these supervised and unsupervised methods.

————————— ◆ —————————

## 1. INTRODUCTION

In recent times, information system security has become a growing concern as computer systems worldwide become increasingly vulnerable or open to harm due to the rapid increase in connectivity and accessibility, which has indirectly resulted in more frequent intrusions, misuses, and attacks. Since most of these problems may be traced or detected through the inspection of patterns of users activities within log and usage files, several intrusion detection systems have been built by manipulating the identified attack and misuse patterns. Many intrusion detection systems use different techniques both anomaly and misuse intrusion detection. The techniques applied in these systems to identify anomalies are varied. While some are based on methods of predicting future patterns of behavior using patterns seen so far [4].

- Hamdan.O.Alanazi – Master Student, Department of Computer System & Technology, University Malaya, Kuala Lumpur, Malaysia
- *Rafidah Md Noor* – Lecturer, Department of Computer Science & Information Technology, University Malaya, Kuala Lumpur, Malaysia
- A. A. Zaidan – PhD Candidate on the Department of Electrical & Computer Engineering , Faculty of Engineering , Multimedia University , Cyberjaya, Malaysia
- B. B. Zaidan – PhD Candidate on the Department of Electrical & Computer Engineering / Faculty of Engineering, Multimedia University, Cyberjaya, Malaysia

Others primarily rely on statistical approaches to establish anomalous behavior. In both cases, observed behavior that does not correspond with expected behavior is aged because an intrusion might be implied [2],[3].

Intrusion Detection attempts to detect computer attacks by inspecting data records observed by processes on the same network. Generally, these attacks are divided into two categories, host-based attacks and network-based attacks. Hostbased attack detection routines normally use system call data from an auditprocess that tracks all system calls made on behalf of each user on a particular machine. These audit processes usually run on each monitored machine [4].

Network-based attack detection routines, meawnhile, usually use network traffic data from a network packet sniffer (e.g., tcpdump). Many computer networks, including the commonly accepted Ethernet (IEEE 802.3) network, use a shared medium for communication. Therefore, the packet sniffer only needs to be on the same shared subnet as the monitored machines[5]

## 2. ARTIFICIAL NEURAL NETWORK ANN-IDS

An Artificial Neural Network (ANN) is comprised of a collection of processing elements that are highly interconnected, and convert a set of inputs to a set of desired outputs. The outcome of the transformation is determined by the traits or characteristics of the elements, and the weights associated with the interconnections among them. By altering the connections between the nodes, the network is able to adapt to the desired outputs [1], [2].

Unlike expert systems, this can provide the user with a definitive answer if the characteristics, which are reviewed, perfectly match those which have been coded in the rule base. Neural network performs an analysis of the information, and presents a probability estimate that the data matches the characteristics, which it has been trained to recognize. While the possibility of a match established by a neural network can be 100%, the precision or accuracy of its decisions entirely depends on the experience the system gains in analyzing examples of the stated problem. Initially, the neural network obtains the experience by training the system to accurately identify preselected examples of the problem. The feedback of the neural network is then assessed and the configuration of the system is improved and perfected until the neural network's analysis of the training data attains a satisfactory level. Apart from the initial training period, the neural network also gains experience over time as it carries out analyses on data related to the problem [7].

## 3. SUPPORT VECTOR MACHINE SVM-IDS

Support Vector Machines [5], or SVMs, are learning machines that plot the training vectors in high dimensional feature space, labeling each vector by its class. SVMs look at the classification problem as a quadratic optimization problem. They combine generalization control with a method to prevent the "curse of dimensionality" by placing an upper bound on the margin between the different classes, making it a practical tool for large and dynamic data sets. The categorization of data by SVMs is done by determining a set of support vectors, which are members of the set of training inputs that outline a hyper plane in feature space. [6]

There are two main reasons for our experimentation with SVMs for intrusion detection. The first is speed because real time performance is of key importance to intrusion detection systems, and any classifier that can potentially outrun neural networks is worth considering. The second reason is scalability: SVMs are relatively insensitive to the number of data points and the classification complexity does not depend on the dimensionality of the feature space [7]

## 4. SELF ORGANIZING MAP SOF-IDS

Unsupervised learning methods using SOM provide a simple and efficient way to classify data sets. To process real-time data for classification, we consider SOMs to be best suited due to their high speed and fast conversion rates, as compared with other learning techniques. In addition to this, SOMs also preserve topological mappings between representations, a feature which is preferred when categorizing normal vs. intrusive behavior for network data. That is, the relationships between senders, obtained sample results statically by collecting different sample network traffic representing normal as well as DoS attack [8].

## 5. FUZZY-IDS

With the fuzzy input sets defined, the next step is to write the rules to identify each type of attack. A collection of fuzzy rules with the same input and output variables is called a fuzzy system. We believe the security administrators can use their expert knowledge to help create a set of rules for each attack.

The rules are created using the fuzzy system editor contained in the Matlab Fuzzy Toolbox. This tool contains a graphical user interface that allows the rule designer to create the member functions for each input or output variable, create the inference relationships between the various member functions, and to examine the control surface for the resulting fuzzy system. It is not expected, however, that the rule designer utterly relies on intuition to create the rules. Visual data mining can assist the rule designer in knowing which data features are most appropriate and relevant in detecting different kinds of attacks [9].

## 6. TYPE OF ATTACKERS

### 6.1 Host and Port Scanning

Attackers often conduct host and port scans as Precursors to other attacks. An intruder will try to establish the existence of hosts on a network or whether a particular service is in use. A host scan is normally characterized by unusual number of Connections to hosts on the network from an uncommon origin. The scans may use a variety of Protocols, and may also utilize an identifier called an SDP to represent a unique link between a source, destination, and a service port [6],[11].

### 6.2 Denial of Service Detection

A common attack scenario is when an attacker overwhelms a target machine with too much data. This chokes the target and inhibits it from performing its intended role. Denial of service (dos) attacks can take a variety of forms, and use different types of Protocols [3]. We developed a representative Fuzzy System for a common dos attack based on ICMP Traffic congestion, and to test the system, we launched an ICMP dos attack called ping flood against a target in a controlled environment, collected the network traces and input the resulting data to the fuzzy system [11].

### 6.3 Unauthorized Servers Detection

Another intrusion detection scenario, which is potentially more damaging than the previous two scenarios, is when an attacker invades a system and install a backdoor or Trojan horse program that can lead to further compromise. Telltale activity that can help identify such intrusions include identifying unusual service ports that are in use on the network, unusual numbers of connections from foreign or unfamiliar hosts, and/or unusual amounts of network traffic load to/from a host on the network [10],[11].

### 6.4 KDD Cup 1999 Data

This was the data set used for The Third International Knowledge Discovery and Data Mining Tools Competition, which was held in conjunction with KDD-99, the Fifth International Conference on Knowledge Discovery and Data Mining. The main objective of the competition was to build a network intrusion detector, a predictive model capable of making a distinction between bad''[1] connections, called intrusions or attacks, and ``good'' normal connections. This database contains a standard set of data to be audited, which include a wide variety of intrusions simulated in a military network environment. Most of the intrusion detection systems since 1999 have been tested and trained on this dataset [10].

## 7. CONCLUSION

In this paper, we have presented two types of AI system, both supervised and unsupervised. While the ANN and SVM present the supervised methods, the SOM and Fuzzy Logic present the unsupervised methods. In this research paper, we can conclude that hybrid-based approaches can overcome problems that appear in the prediction of the IDS. Recent researches have invented many hybrid methods using KNN, HMM, Naïve Bayesian, ANN, Fuzzy Logic, SVM, SOM, etc. All these methods can be simulated using the Matlab and KDD99 dataset.

### ACKNOWLEDGEMENT

The author would like to thank ALLAH S.W.T for giving him the resources and ability to write this paper. The author would also like to thank his supervisor, Miss Rafidah Md Noor, for the unlimited support and guidance rendered during the preparation of this research paper. Last but not least, the author would like to express his gratitude to all his friends, and individuals, who have directly and indirectly assisted in the completion of his work.

**Hamdan Al-Anazi**: has obtained his bachelor dgree from "King Suad University", Riyadh, Saudi Arabia. He worked as a lecturer at Health College in the Ministry of Health in Saudi Arabia, then he worked as a lecturer at King Saud University in the computer department. Currently he is Master candidate at faculty of Computer Science & Information Technology at University of Malaya in Kuala Lumpur, Malaysia. His research interest on Information Security, cryptography, steganography and digital watermarking, He has contributed to many papers some of them still under reviewer.

**Rafidah Md Noor:** She obtained here Bachelor degree in information technology (Hons), uum, Sintok, Kedah, 1998 followed by master of Science (computer science), UTM, skudai, johore, 2000, Here interest area are Network Mobility (NEMO) Basic Support protocol is defined by IETF in RFC3963. The group is actively work on issues such as routing optimization and multihoming but not specifically in Quality of Service (QoS). In infolab21, we have agreed to work on QoS provisioning for my research work. During my first year study as MPhil student, she has worked on dynamic QoS provisioning proposal and NEMO BS protocol implementation. The implementation has been carried out for 4 months and 2 months for data analysis. A dynamic QoS provisioning for Network Mobility is concerned on resource reservation at the mobile nodes i.e. Mobile Router and Mobile Network Nodes. We have proposed 4-tier model which covered from the application layer until the data link layer. Our novel idea is a contract-basis resource reservation where we adapted the concept of different QoS classes in DiffServ technology. To implement the idea, we considered three different real-life scenarios from a personal transport (Car) till a public transport (Train). Currently, she is PhD Student Computing Department Infolab21 South Drive Lancaster University LA1 4WA Lancaster United Kingdom and she is a lecturer, Department of Computer System &, Technology/University of Malaya / Kuala Lumpur/Malaysia.

**Bilal Bahaa Zaidan:** He obtained his bachelor degree in Mathematics and Computer Application from Saddam University/Baghdad followed by master in data communication and computer network from University of Malaya. He led or member for many funded research projects and He has published more than 50 papers at various international and national conferences and journals, His interest area are Information security (Steganography and Digital watermarking), Network Security (Encryption Methods) , Image Processing (Skin Detector), Pattern Recognition , Machine Learning (Neural Network, Fuzzy Logic and Bayesian) Methods and Text Mining and Video Mining. .Currently, he is PhD Candidate on the Department of Electrical & Computer Engineering / Faculty of Engineering / Multimedia University / Cyberjaya, Malaysia. He is members IAENG, CSTA, WASET, and IACSIT. He is reviewer in the (IJSIS, IJCSNS, IJCSN, IJCSE and IJCIIS).

**Aos Alaa Zaidan**: He obtained his 1st Class Bachelor degree in Computer Engineering from university of Technology / Baghdad followed by master in data communication and computer network from University of Malaya. He led or member for many funded research projects and He has published more than 50 papers at various international and national conferences and journals, His interest area are Information security (Steganography and Digital watermarking), Network Security (Encryption Methods) , Image Processing (Skin Detector), Pattern Recognition , Machine Learning (Neural Network, Fuzzy Logic and Bayesian) Methods and Text Mining and Video Mining. .Currently, he is PhD Candidate on the Department of Electrical & Computer Engineering / Faculty of Engineering / Multimedia University / Cyberjaya, Malaysia. He is members IAENG, CSTA, WASET, and IACSIT. He is reviewer in the (IJSIS, IJCSNS, IJCSN, IJCSE and IJCIIS).